# Hierarchical Approach for Key Management in Mobile Ad hoc Networks

Renuka A.

Dept. of Computer Science and Engg.
Manipal Institute of Technology
Manipal-576104-India
renuka.prabhu@manipal.edu

Dr. K.C.Shet

Dept. of Computer Engg.
National  Institute of Technology Karnataka
Surathkal, P.O.Srinivasanagar-575025
kcshet@rediffmail.com

*Abstract*—**Mobile Ad-hoc Network (MANET) is a collection of autonomous nodes or terminals which communicate with each other by forming a multi-hop radio network and maintaining connectivity in a decentralized manner. The conventional security solutions to provide key management through accessing trusted authorities or centralized servers are infeasible for this new environment since mobile ad hoc networks are characterized by the absence of any infrastructure, frequent mobility, and wireless links. We propose a hierarchical group key management scheme that is hierarchical and fully distributed with no central authority and uses a simple rekeying procedure which is suitable for large and high mobility mobile ad hoc networks. The rekeying procedure requires only one round in our scheme and Chinese Remainder Theorem Diffie Hellman Group Diffie Hellmann and Burmester and Desmedt it is a constant 3 whereas in other schemes such as Distributed Logical Key Hierarchy and Distributed One Way Function Trees, it depends on the number of members. We reduce the energy consumption during communication of the keying materials by reducing the number of bits in the rekeying message. We show through analysis and simulations that our scheme has less computation, communication and energy consumption compared to the existing schemes.**

## I. INTRODUCTION

A mobile ad hoc network (MANET) is a collection of autonomous nodes that communicate with each other, most frequently using a multi-hop wireless network. Nodes do not necessarily know each other and come together to form an ad hoc group for some specific purpose. Key distribution systems usually require a trusted third party that acts as a mediator between nodes of the network. Ad hoc networks typically do not have an online trusted authority but there may be an off line one that is used during system initialization.

Group key establishment means that multiple parties want to create a common secret to be used to exchange information securely. Without relying on a central trusted entity, two people who do not previously share a common secret can create one based on the party Diffie Hellman (DH) protocol. The 2-party Diffie Hellman protocol can be extended to a generalized version of n-party DH. Furthermore, group key management also needs to address the security issue related to membership changes. The modification of membership requires refreshment of the group key. This can be done either by periodic rekeying or updating right after member change. The change of group key ensures backward and forward security. With frequently changing group memberships, recent researches began to pay more attention on the efficiency of group key update. Recently, collaborative and group-oriented applications in MANETs have been an active research area. Obviously, group key management is a central building block in securing group communications in MANETs. However, group key management for large and dynamic groups in MANETs is a difficult problem because of the requirement of scalability and security under the restrictions of nodes' available resources and unpredictable mobility.

We propose a distributed group key management approach wherein there is no central authority and the users themselves arrive at a group key through simple computations. In large and high mobility mobile ad hoc networks, it is not possible to use a single group key for the entire network because of the enormous cost of computation and communication in rekeying. So, we logically divide the entire network into a number of groups headed by a group leader and each group is divided into subgroups called clusters headed by the cluster head. Though the term group leaders and cluster heads are used these nodes are no different from the other nodes, except for playing the assigned roles during the initialization phase and inter group and inter cluster communication. After initialization phase, within any cluster, any member can initiate the rekeying process and the burden on the cluster head is reduced. The transmission power and memory of the cluster head and the group leaders is same as other members. The members within the cluster communicate with the help of a group key. Inter cluster communication take place with the help of gate way nodes if the nodes are in the adjacent clusters and through the cluster heads if the are in far off clusters.. Inter group communication is routed through the group leaders. Each member also carries a public key, private key pair used to encrypt the rekeying messages exchanged. This ensures that the forward secrecy is preserved.

The rest of the paper is organized as follows. Section II focuses on the related work in this field. The proposed scheme is presented in Section III. Performance analysis of the scheme is discussed in Section IV. Experimental Results and Conclusion are given in Section V and Section VI respectively.

## II. RELATED WORK

Key management is a basic part of any secure communication. Most cryptosystems rely on some underlying secure, robust, and efficient key management system. Group key establishment means that multiple parties want to create a common secret to be used to exchange information securely. Secure group communication (SGC) is defined as the process by which members in a group can securely communicate with each other and the information being shared is inaccessible to anybody outside the group. In such a scenario, a group key is established among all the participating members and this key is used to encrypt all the messages destined to the group. As a result, only the group members can decrypt the messages. The group key management protocols are typically classified in four categories: centralized group key distribution (CGKD), de-centralized group key management (DGKM), distributed/contributory group key agreement (CGKA), and distributed group key distribution (DGKD).

In CGKD, there exists a central entity (i.e. a group controller (GC)) which is responsible for generating, distributing, and updating the group key. The most famous CGKD scheme is the key tree scheme (also called Logical Key Hierarchy (LKH) proposed in [1] is based on the tree structure with each user (group participant) corresponding to a leaf and the group initiator as the root node. The tree structure significantly reduces the number of broadcast messages and storage space for both the group controller and group members. Each leaf node shares a pairwise key with the root node as well as a set of intermediate keys from it to the root. One Way Function (OFT) is another centralized group key management scheme proposed in [2].similar to LKH. However, all keys in the OFT scheme are functionally related according to a one-way hash function

The DGKM approach involves splitting a large group into small subgroups. Each subgroup has a subgroup controller which is responsible for the key management of its subgroup. The first DGKM scheme to appear was IOLUS [3]. The CGKA schemes involve the participation by all members of a group towards key management. Such schemes are characterized by the absence of the GC. The group key in such schemes is a function of the secret shares contributed by the members.Typical CGKA schemes include binary tree based ones [4] and n-party Diffie-Hellman key agreement [5, 6]. Tree Based Group Diffie Hellman (TGDH) is a group key management scheme proposed in [4]. The basic idea is to combine the efficiency of the tree structure with the contributory feature of DH. The DGKD scheme, proposed in [7], eliminates the need for a trusted central authority and introduces the concepts of sponsors and co distributors. All group members have the same capability and are equally trusted. Also, they have equal responsibility, i.e. any group member could be a potential sponsor of other members or a co-distributor. Whenever a member joins or leaves the group, the member's sponsor initiates the rekeying process. The sponsor generates the necessary keys and securely distributes the keys to co-distributors respectively. The co distributors then distribute in parallel, corresponding keys to corresponding members. In addition to the above four typical classes of key management schemes, there are some other forms of key management schemes such as hierarchy and cluster based ones [6, 8]. A contributory group key agreement scheme is most appropriate for SGC in this kind of environment.

Several group key management schemes have been proposed for SGC in wireless networks [9, 10]. In Simple and Efficient Group Key (SEGK) management scheme for MANETs proposed in [11] group members compute the group key in a distributed manner. Also, a new approach was developed in [12] called BALADE, based on a sequential multi-sources model, and takes into account both localization and mobility of nodes, while optimizing energy and bandwidth consumptions. Most of these schemes involve complex operations which is not suitable for large and high mobility networks. In Group Diffie-Hellman, the group agrees on a pair of primes and starts calculating in a distributive fashion the intermediate values. The setup time is linear since all members must contribute to generating the group key. Therefore, the size of the message increases as the sequence is reaching the last members and more intermediate values are necessary. With that, the number of exponential operations also increases. Therefore this method is not suitable for large networks. Moreover the computational burden is high since it involves a lot of exponentiations.

Another approach using logical key hierarchy in a distributed fashion was proposed in [13] called Distributed One-way Function Tree (D-OWT) This protocol uses the one-way function tree. A member is responsible for generating its own key and sending the blinded version of this key to its sibling. Reference [14] also uses a logical key hierarchy to minimize the number of key held by group members called Diffie–Hellman Logical Key Hierarchy. The difference here is that group members generate the keys in the upper levels using the Diffie–Hellman algorithm rather than using a one-way function. In Chinese Remainder Theorem Diffie-Hellman (CRTDH) [15] each member computes the group key as the XOR operation of certain values computed. This requires that the members agree on two large primes. CRTDH is impractical in terms of efficiency and security, such as low efficiency, possibly a small key, and possessing the same Least Common Multiple (LCM). However this CRTDH scheme was modified in [16] wherein the evaluation of the LCM was eliminated and other steps were modified slightly, so that a large value for the key is obtained. In both these methods, whenever membership changes occur, the new group key is derived from the old group key as the XOR function of the old group key and the value derived from the Chinese Remainder Theorem values broadcast by one of its members. Since it is possible for the leaving member to obtain this message, and hence deduce the new group key backward secrecy is not preserved.

In this paper, we propose a distributed approach in which members contribute to the generation of group key by sending the hash of a random number during initialization phase within the cluster. They regenerate the group key themselves by obtaining the rekeying message from one of its members during rekeying phase or whenever membership changes occur. In a group the group key used for communication among the cluster heads is generated by the group leader and transmitted securely to the other clusterheads. The same procedure is used to agree on a common key among the group leaders wherein the network head generates the key and passes on to the other group leaders. Symmetric key is used for communication between the members of a cluster and asymmetric key cryptography for distributing the rekeying messages to the members of the cluster.

### III. PROPSED SCHEME

#### A. System model

The entire set of nodes is divided into a number of groups and the number of nodes within a group is further subdivided into subsets called clusters. Each group is headed by a group leader and a cluster by the cluster head. The layout of the network is as shown in Fig.1. One of the nodes in the cluster is head. A set of eight such clusters form a group and each group is headed by a group leader. The cluster head is similar to the nodes in the network. The nodes within a cluster are also the physical neighbors. The nodes within a cluster use contributory key agreement. Each node within a cluster contributes his share in arriving at the group key. Whenever membership changes occur, the adjacent node initiates the rekeying operation thereby reducing the burden on the cluster head. The group leader chooses a random key to be used for encrypting messages exchanged between the cluster heads and the network head sends the key to the group leaders that is used for communication among the group leaders. The hierarchical arrangement of the network is shown in Fig.2.

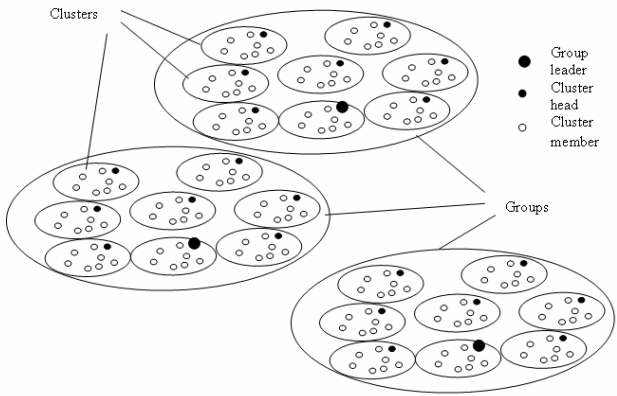

Figure 1. Network Layout

The key management system consists of two phases
(i) Initialization
(ii) Group Key Agreement

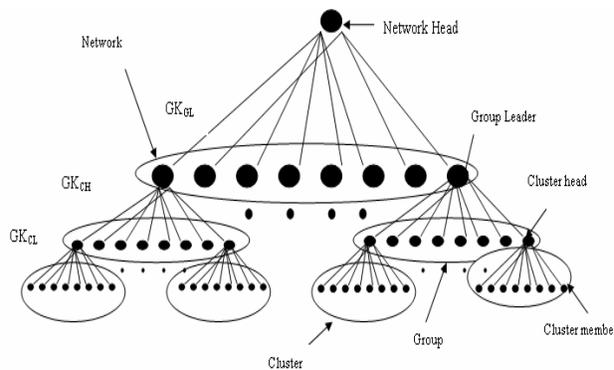

Figure 2. Hierarchical layout

### B. Initialization

Step 1: After deployment, the nodes broadcast their id value to their neighbors along with the HELLO message.
Step 2: When all the nodes have discovered their neighbors, they exchange information about the number of one hop neighbors. The node which has maximum one hop neighbors is selected as the cluster head. Other nodes become members of the cluster or local nodes. The nodes update the status values accordingly.
Step 3: The cluster head broadcasts the message "I am cluster head" so as to know its members.
Step 4: The members reply with the message "I am member" and in this way clusters are formed in the network.
Step 5: If a node receives more than one "I am cluster head" messages, it becomes Gateway which acts as a mediator between two clusters.
In this manner clusters are formed in the network.
The cluster heads broadcast the message, "Are there any cluster heads" so as to know each other. The cluster head with the smallest id is selected as the leader of the cluster heads which is representative of the group called the group leader. The group leaders establish communication with other group leaders in a similar manner and one among the group leaders is selected as the leader for the entire network. The entire network is hierarchical in nature and the following hierarchy is observed
network→group→cluster→cluster members

### C. Group Key Agreement within a cluster

Step 1: Each member broadcasts the public key along with its id to all other members of the cluster along with the certificate for authentication.
Step 2: The members of the cluster generate the group key in a distributive manner. Each member generates a random number and sends the hash of this number to the other members encrypted with the public key of the individual members, so that the remaining members can decrypt the message with their respective private key.
Step 3: Each member concatenates the hash values of the received members in the ascending order of the ids and mixes it using a one way hash function on the concatenated string. This is the group key used for that cluster.
Let $HR_i$ be the hash of the random number generated by node i and GK denote the group key then

$GK = f(HR_1, HR_2, HR_3, \ldots\ldots HR_n)$

where

$HR_i$ = hash(Random number i)

f is a one way function and
hash is secure hash function such as SHA1.
All the members now possess a copy of the same key as same operations are performed by all the nodes.

### D. Inter cluster group key agreement

The gateway node initiates communication with the neighboring node belonging to another cluster and mutually agrees on a key to be used for inter cluster communication between the two clusters. Any node belonging to one cluster can communicate with any other node in another cluster through this node as the intermediary. In this way adjacent clusters agree on group key. A set of eight clusters form a group. The cluster heads of each of these clusters mutually agree on a group key to be used for communication among the clusterheads within a group in the similar manner. This key is different from the key used within the cluster. Going one level above in the hierarchy, a number of groups can be combined headed by a group leader and the group leaders agree on a group key to be used for communication among the group leaders which aids in intergroup communication.

Even though we have considered that the network is divided into eight groups, each group consisting of eight clusters and each cluster consisting of eight members, it need not be constant. It may vary and this number does not change the manner in which group key is derived. This is assumed so that it gives the hierarchical appearance in the form of a fractal tree.

*E.  Network Dynamics*

The mobile ad hoc network is dynamic in nature. Many nodes may join or leave the network. In such cases, a good key management system should ensure that backward and forward secrecy is preserved.

*1) Member join:*

When a new member joins, it initiates communication with the neighbouring node. After initial authentication, this node initiates the rekeying operations for generating a new key for the cluster. The rekeying operation is as follows.

new node $\rightarrow$ adjacent node : {authentication}

adjacent node $\rightarrow$ new node :{acknowledge}

adjacent node $\rightarrow$ all nodes:{rekeying message}$k_{(old\ cluster\ key)}$

The neighboring node broadcasts two random numbers that are mixed together using a hashing function and is inserted at a random position in the old group key, the position being specified by the first random number. The two random numbers are sent in a single message, so that any transmission loss may not result in wrong key being generated.  Let the two bit strings be

I Random no. = 00100010

II Random no. = 10110111

Suppose the result of mixing function is 11010110

and the previous group key is

1001010001010101000111000011110000011000010000001

The new group key is

1001010001010101000111000011110000011001**1010110**010000001

Since all members know the old group key they can compute the new group key.

This new group key is transmitted to the new member by the adjacent node in a secure manner.

*2) Member Leave*

*a)  When Cluster Member leaves*

When a member leaves the group key of the cluster to which it belongs must be changed. This is changed in the similar manner as described above. The leaving member informs the neighboring node which in turn informs the other nodes about the leaving member. It also generates two random numbers and sends it securely to the other members which generate the group key.

leaving  node $\rightarrow$ adjacent node : {leaving message}

adjacent node $\rightarrow$ leaving node :{acknowledge}

adjacent node $\rightarrow$ each  node $_i$ :{rekeying message}$pk_i$

*b)  When a gateway node leaves*

When a gateway node leaves the network, it delegates the role of the gateway to the adjacent node. In this case, the group key of both the clusters with which this node is associated need to be changed. When the gateway node moves into one of the clusters only the group key of the other cluster has to be changed.

leaving  gateway node  $\rightarrow$ adjacent node : {leaving message + other messages for delegating its role}

adjacent  node $\rightarrow$ leaving  gateway  node :{acknowledge}

adjacent node $\rightarrow$ each node $_i$ in cluster1:{rekeying message}$pk_i$

adjacent node $\rightarrow$ each  node $_j$ in cluster2:{rekeying message}$pk_j$

*c)  When the cluster head leaves*

When the cluster head leaves the group key used for communication among the cluster heads need to be changed. Also, the group key used within the cluster has to be changed. This cluster head informs the adjacent cluster head about its desire to leave the network which initiates the rekeying procedure. The adjacent cluster head generates two random numbers and sends it to the other cluster heads in a secure manner.

leaving  cluster head $\rightarrow$ adjacent cluster head : {leaving message}

adjacent node $\rightarrow$ leaving node :{acknowledge}

adjacent node $\rightarrow$ each  node $_i$ :{rekeying message}$pk_i$

leaving  clusterhead $\rightarrow$ adjacent clusterhead : {leaving message + other messages for delegating its role}

adjacent node $\rightarrow$ leaving clusterhead :{acknowledge}

adjacent node $\rightarrow$ each  cluster head$_i$  :{rekeying message}$pk_i$

The group key of the clusterheads is obtained by taking the product of the two random numbers, inserting at the position of indicated by the first number and removing the initial bits old group key of the clusterheads and removing the bits equal to the number of bits in the product from the old group key.

Suppose

I  Random no. = 00101101

II Random no. = 00111111

The product of the two numbers is00 00010101000110

Suppose the old group key is

1001010001010101000111000011110000011000100000010110

The new group key is

0001110000111100000110001000**000010101000110**000100110. Thus the cluster heads compute the group key after rekeying operation. This is the new group key for clusterheads within a group.Even the group key used for intra cluster communication in that particular cluster needs to be changed. This is changed in the manner described above for rekeying within the clauster.

*d)* Whenever the group leader leaves

Whenever the group leader leaves all the three keys should be changed. These are

(i) group key among the group leaders

(ii)group key among the clusterheads and

(iii) group key within the cluster

leaving group leader→adjacent group leader:{leaving message + other messages for delegating its role }

adjacent group leader→leaving node :{acknowledge}

adjacent group leader →each group leader$_i$ :{rekeying message}pk$_i$

leaving group leader→adjacent node : {leaving message}

adjacent node→leaving group leader :{acknowledge}

adjacent node →each node $_i$ in that cluster:{rekeying message}pk$_i$

leaving group leader →adjacent clusterhead : {leaving message + other messages for delegating its role}

adjacent clusterhead →leaving clusterhead :{acknowledge}

adjacent node →each cluster head$_i$ :{rekeying message}pk$_i$

leaving node →adjacent node : {leaving message}

adjacent node →leaving node :{acknowledge}

adjacent node →each node $_i$ :{rekeying message}pk$_i$

The first two group keys are changed in the manners described above. To change the group key of the group leaders, the leaving group leader delegates the role of the group leader to another cluster head in the same group and informs it to the other group leaders about this change. The adjacent group leader initiates the rekeying operation. It generates two random numbers, and sends it the other group leaders. The group leaders divide the old group key into blocks of size which is the same as the number of bits in the random number, perform the exclusive OR of the random number and the blocks of the old group key and concatenate the result to arrive at the new group key.

Suppose the Random no. is 00100010 and the old group key is

1001010001010101000111000011110000011000100000010110101

Dividing the group key into 8 bit blocks (size of the random number), we get,

10010100 01010101 00011100 00111100 00011000 10000001 00110101

Performing the XOR operation and concatenating as shown,

00100010 XOR 10010100 ‖ 00100010 XOR 01010101 ‖ 00100010 XOR 00011100 ‖ 00100010 XOR 00111100 ‖ 00100010 XOR 00011000 ‖ 00100010 XOR 10000001 ‖00100010 XOR 00110101

the following group key is obtained

1011011001110111001111100001110001110101010001100010111

This is the new group key of the group leaders.

*F. Communication Protocol*

The nodes within the cluster communicate using the intra cluster group key. The communication between intra group and inter cluster nodes takes place through the gateway node, if they belong to adjacent clusters and through the cluster heads if the are in far off clusters.
Sourcenode→gateway node→Destination node --- For adjacent clusters
Sourcenode→clusterhead(source)→clusterhead(destination)→Destination node ---For far away clusters

For adjacent clusters

     GK$_{CL1}$      GK$_{CL2}$
Source node ------------>Gateway node ---------->
Destination node
For nodes in far off clusters

     GK$_{CL1}$      GK$_{CH}$
Source node ------------>Cluster head1 ----------> Cluster
     GK$_{CL2}$
head1 ------------→ Destination node

The inter group communication is through corresponding cluster heads and the group leaders as shown
Source node→cluster head (source)→group leader (source) →group leader (destination) →cluster head (destination) →Destination node.

     GK$_{CL1}$      GK$_{CH1}$
Source node ------------>Cluster head1 ----------> Group
     GK$_{GR}$      GK$_{CH2}$
Leader1 ------------→Group Leader2-------------→Cluster
     GK$_{CL2}$
head2 ------------→ Destination node

## IV. PERFORMANCE ANALYSIS

### A. Cost Analysis

We compute communication cost of our scheme under various situations and for different network organizations. We also compare the communication cost of rekeying for various schemes. Some schemes such as GD-H use 1024 bit message for rekeying whereas our sehme uses a 32 bit meaasge and therefore the energy required for rekeying is very less. This is very important in energy constrained mobile ad hoc networks.

Let us denote

N= Network size
M=Group Size
P=Cluster Size
G=No. of groups
CH=Cluster Head
CL=Cluster member
GL =Group leader

#### 1) Member joins

When a new member joins, the public key of the new member is broadcast to all old members encrypted with the old group key. Suppose the average number of members in a cluster is P, two 16 bit numbers or a message of 32 bits is transmitted to all the existing members encrypted with the old key. This scheme requires one round and 1 broadcast message . The group keys of other clusters need not be changed.

#### 2) Member leaves

When a node leaves, there are three cases
(i) The cluster member leaves
(ii)The cluster head leaves
(iii)The gateway node leaves
(iv)The group leader leaves

##### a) When the cluster member leaves

The random numbers are sent to the existing members encrypted with their respective public keys and unicast to the existing members. Therefore this requires one round and P-1 unicast messages.

##### b) When the cluster head leaves

The rekeying is similar to member leave within the cluster i.e P-1 unicast messages and M-1 messages among the cluster heads for changing the cluster head key.

##### c) When gateway node leaves

The group key of both the cluster with which it is associated have to change the group keys. Therefore, this requires one round in each cluster and M-1 unicast messages in each cluster that is a total of 2 (P-1) messages.

##### d) When the group leader leaves

The group key of the group leaders , the group key of the cluster heads and also the cluster key of the cluster need to be changed. This requires one round and G-1 unicast messages among the group leaders, M-1 unicast messages among the group leaders and P-1 messages within the cluster.

Table II gives the communication cost of rekeying for various schemes. In our scheme, the entire network is divided into a number of groups which in turn is divided into a number of clusters, wherein each cluster consists of members. When a member leaves, in the non hierarchical scheme, the key of the entire network needs to be changed. But in hierarchical scheme, it is just sufficient if the group key of the cluster to which it belongs is changed. The hierarchical scheme reduces the number of rekeying messages transmitted and this is shown in Table I. The communication between far off nodes (nodes in different groups) has to undergo 5 encryptions and decryptions whereas in non hierarchical schemes it is only one. In very large networks, this is tolerable compared to the enormous rekeying messages that need to be transmitted whenever membership changes occur. From this table we observe that the rekeying procedure requires only one round in our scheme and CRTDH and modified CRTDH, in GD-H and BD it is a constant 3 whereas in other schemes such as D-LKH and D-OFT, it depends on the number of members. Regarding the number of messages sent, BD method involves 2N broadcast messages and no unicast messages, whereas in our technique, the number of unicast messages is N-1. We also observe that CRTDH has the least communication cost among all the methods, but it does not provide forward secrecy because the rekeying message is broadcast and even the leaving member can derive the new group key. Moreover, in our scheme the rekeying message is only 32 bits wide and thus the communication overhead is greatly reduced.

TABLE I. NO. OF REKEYING MESSAGES FOR DIFFERENT NETWORK SIZES

| Network Organization | No. of nodes that receive rekeying messages (Our scheme) | | | Non-hierarchical scheme |
|---|---|---|---|---|
| | CL leaves or CL joins | CH leaves | GL leaves | |
| N=256 M=8 P=16 G=2 | Join -15 (Broadcast) Leave-15 | 32 | 34 | 256 |
| N=256 M=4 P=16 G =4 | Join -16 Leave-15 | 32 | 36 | 256 |
| N=256 M=4 P=8 G =8 | Join -8 Leave-7 | 40 | 44 | 256 |
| N=256 M=4 P=4 G =16 | Join -4 Leave-3 | 68 | 72 | 256 |
| N=256 M=2 P=4 G=32 | Join -4 Leave-3 | 68 | 70 | 256 |

TABLE II.    COMMUNICATION COST OF REKEYING

| Scheme | No. of rounds | No. of messages | |
|---|---|---|---|
| | | *Broadcast* | *Unicast* |
| Burmester and Desmedt(BD) | 3 | 2N | 0 |
| Group-Diffie Hellman(GDH) | N | N | N-1 |
| Distributed Logical Key Hierarchy (D-LKH) | 3 | 1 | N |
| Distributed One Way Function Trees(D-OFT) | $Log_2$ N | 0 | $2Log_2$ N |
| CRTDH | 1 | 1 | |
| Modified CRTDH | 1 | 1 | |
| Our scheme(join) | 1 | 1 | 0 |
| Our scheme CL (leave) | 1 | 0 | P-1 |
| Gateway leave | 1 | 0 | 2(P-1) |
| CH leave | 1 | 0 | M+P-2 |
| GL leave | 1 | 0 | G+P+M-3 |

Let
Exp=Exponential operation
D=Decryption operation
OWF=One Way Function
X=Exclusive OR operation
CRT=Chinese Remainder Theorem method for solving congruence relation
i= node id
M= Cluster size

TABLE III.    COMPUTATIONAL COMPLEXITY

| Scheme | During Set up phase | | During rekey |
|---|---|---|---|
| | *Cluster head* | *Members* | |
| Burmester and Desmedt | (M+1)Exp | ----- | (M+1)Exp |
| Group-Diffie Hellman | (i+1)Exp | ------ | (i+1)Exp |
| Distributed Logical Key Hierarchy | $Log_2$(MExp) | $Log_2$MD | $Log_2$MD |
| Distributed One Way Function Trees | $(Log_2$M+ 1) Exp | ------ | $(Log_2$M + 1)Exp |
| CRTDH | ----- | LCM(M-1) + (M-1) X +MExp + CRT | LCM+X+CRT →leader CRT+X→members |
| Modified CRTDH | ------ | (M-1) X +MExp +CRT | X+CRT |
| Our scheme | Sort+ OWF | Sort+ OWF | D+OWF Multiplication( CH leave) XOR(GL leave) |

## V.    EXPERIMENTAL RESULTS

The simulations are performed using Network Simulator (NS-2.32) [17], particularly popular in the ad hoc networking community. The MAC layer protocol IEEE 802.11 is used in all simulations. The Ad Hoc On-demand Distance Vector (AODV) routing protocol is chosen for the simulations. Every simulation run is 500 seconds long. The simulation is carried out using different number of nodes. The simulation parameters are shown in Table III.

The experiment is conducted with different mobility patterns generated using the setdest tool of ns2. These are stationary nodes located at random positions, nodes moving to random destinations with speeds varying between 0 and a maximum of 5m/s, 10m/s and 20m/s. The random waypoint mobility model is used in which the nodes move to a randomly selected position with the speed varying between 0 and maximum speed, pauses for a specified pause time and again starts moving with the same speed to a new destination. The pause time is set to 200 secs. Different message sizes of 16, 32, 48, 64, 128, 152, 180, 200, 256, 512 and 1024 bits are used. We observed that in all the four scenarios the energy consumed by the node increases as the message size increases. This is depicted in Fig.3. Since the nodes in a mobile ad hoc network communicate in a hop by hop manner, the energy consumed by all the nodes is not the same, even though same number of messages are sent and received by the nodes. This is clearly visible from the graphs. From the graph we observe that the energy consumed is less for a speed of 10m/s. This may be due to the fact that the movement brings the nodes closer to each other which reduces the relaying of the messages. The energy shown is inclusive of the energy for forwarding the message by the intermediate node.

TABLE IV.    SIMULATION PARAMETERS

| Parameters | Values |
|---|---|
| Simulation time | 1000 sec |
| Topology size | 500m X 500m |
| Initial energy | 100 Joules |
| Transmitter Power | 0.4W |
| Receiver Power | 0.3W |
| Node mobility | Max. speed 0m/s,5m/s, 10m/s, 20m/s |
| Routing Protocol | AODV |
| Traffic type | CBR, Message |
| MAC | IEEE 802.11 |
| Mobility model | Random Waypoint |
| Max. no. of packets | 10000 |
| Pause time | 200sec |

In the next experiment, we varied the cluster size and observed the effect of the cluster size on the average energy consumed by the nodes for communicating the rekeying messages. In this setup one node sends a message to every other node in the cluster. For P nodes, P-1 messages are exchanged. This is indicated in Fig 4 for the mobility pattern of max. speed 20m/s. We observe that the energy consumed by the nodes increases as the network size increases and this is true with message sizes also.

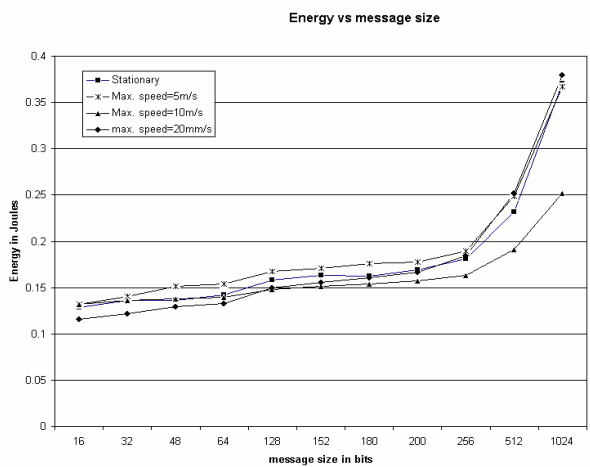

Figure 3.   Average energy consumed by the nodes for various message sizes for cluster size of 8 nodes

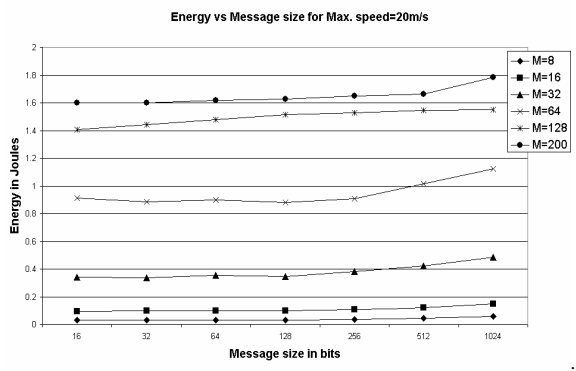

Figure 4.   Average energy consumed by the nodes vs. message size for different cluster sizes with mobility pattern of max. speed=20m/s

## VI. CONCLUSION

We proposed a hierarchical scheme scheme for group key management that does not rely on a centralized authority for regenerating a new group key. Any node can initiate the process of rekeying and so the energy depletion of any one particular node is eliminated unlike the centralized schemes. Our approach satisfies most of the security attributes of a key management system. The communication and computational overhead is small in our scheme compared with other distributed schemes. The energy saving is approximately 41% for 8 nodes and 15% for 200 nodes when the message size is reduced from 1024 to 16 bits. This indicates that small message size and small cluster size is most suitable for energy limited mobile ad hoc networks. A small cluster size increases the overhead of inter cluster communication since it needs more encryptions and decryptions whereas a large cluster size increases the communication cost of rekeying. An optimal value is chosen based on the application. As a future work, instead of unicasting the rekeying messages, broadcasting may be done that will reduce the number of messages sent through the network.Since the leaving member should not have access to this information, doing this in a secure manner is a challenging task.